\documentstyle[twoside,fleqn,bezier]{article}

\def\noi{\noindent}

\makeatletter

\renewcommand{\section}{\@startsection{section}{1}{0pt}%
        {-3.5ex plus -1ex minus -.2ex}{2.3ex plus .2ex}%
        {\large\bf\protect\raggedright}}

\renewcommand{\subsection}{\@startsection{subsection}{2}{0pt}%
        {-3ex plus -1ex minus -.2ex}{1.4ex plus .2ex}%
        {\normalsize\bf\protect\raggedright}}

\renewcommand{\thesubsubsection}%
        {\arabic{section}.\arabic{subsection}.\arabic{subsubsection}.}
\renewcommand{\@oddhead}{\raisebox{0pt}[\headheight][0pt]{%
   \vbox{\hbox to\textwidth{\rightmark \hfil \rm \thepage \strut}\hrule}}}
\renewcommand{\@evenhead}{\raisebox{0pt}[\headheight][0pt]{%
   \vbox{\hbox to\textwidth{\thepage \hfil \leftmark \strut}\hrule}}}
\newcommand{\heads}[2]{\markboth{\protect\small\it #1}{\protect\small\it #2}}
\newcommand{\Acknow}[1]{\subsection*{Acknowledgement} #1}

\def\ncirc#1#2{%
  \font\newcirc=lcircle10 scaled #2
  \font\newcircw=lcirclew10 scaled #2
     \def\thinlines{\let\@linefnt\tenln \let\@circlefnt\newcirc
        \@wholewidth\fontdimen8\tenln \@halfwidth .5\@wholewidth}
  \def\thicklines{\let\@linefnt\tenlnw \let\@circlefnt\newcircw
       \@wholewidth\fontdimen8\tenlnw \@halfwidth .5\@wholewidth}
 \thinlines
 \circle{#1}}

\makeatother

\newcommand{\Title}[1]{\noindent {\Large #1} \\}

\newcommand{\Abstract}[1]{\vskip 2mm \begin{center}
     \parbox{16.4cm}{\small\noindent #1} \end{center}\bigskip}
\newcommand{\foom}[1]{\protect\footnotemark[#1]}

\newcommand{\email}[2]{\footnotetext[#1]{e-mail: #2}}

\newcommand{\sect}[1]{Sec.\,#1}

\def\nqq{\hspace{-2em}}
\def\nhq{\hspace{-0.5em}}

\def\cm{\hspace{1cm}}
\def\inch{\hspace{1in}}

\newcommand{\sequ}[1]{\setcounter{equation}{#1}}
\newcommand{\Eq}[1]{Eq.\,(\ref{#1})}

\def\eq{Eq.\,}

\def\beq{\begin{equation}}
\def\eeq{\end{equation}}
\def\bear{\begin{eqnarray}}
\def\al{&\nhq}
\def\lal{&&\nqq {}}               
\def\bearr{\begin{eqnarray} \lal}
\def\ear{\end{eqnarray}}
\def\earn{\nonumber \end{eqnarray}}
\def\dst{\displaystyle}
\def\tst{\textstyle}

\def\nn{\nonumber\\ {}}

\def\nnn{\nonumber\\ \lal }

\def\yy{\\[5pt]}

\def\eql{\al =\al}

\def\eqdef{\stackrel{\rm def}{=}}
\def\e{{\,\rm e}}
\def\d{\partial}

\def\sign{{\,\rm sign\,}}

\def\dim{{\,\rm dim\,}}
\def\const{{\rm const}}
\def\Half{{\dst\frac{1}{2}}}
\def\half{{\tst\frac{1}{2}}}
\def\then{\ \Rightarrow\ }

\newcommand{\aver}[1]{\langle \, #1 \, \rangle \mathstrut}
\def\DAL{\raisebox{-1.6pt}{\large $\Box$}\,}
\newcommand{\vars}[1]{\left\{\begin{array}{ll}#1\end{array}\right.}

\def\intl{\int\limits}

\renewcommand{\theequation}{\thesection\arabic{equation}}
\catcode`\@=11 \@addtoreset{equation}{section}\catcode`\@=12
\newcommand{\eps}{\varepsilon}
\newcommand{\sq}[1]{\sqrt{|#1|}}
\def\M{{\cal M}}
\def\oG{{\overline G}}
\def\oc{{\overline c}}
\def\sumn{\sum_{i=1}^{n}}
\def\sumo{\sum_{I\in \Omega_*}}
\def\ubr{\underbrace}
\def\ss{{\scriptstyle s}}
\newcommand{\R}{{\sf R\hspace*{-0.9ex}\rule{0.1ex}{1.5ex}\hspace*{0.9ex}}}

\newcommand{\Picture}[6]{
	\begin{figure}  \unitlength=1mm
	\begin{picture}(82.5,#1)
		\put(0,0){\line(0,1){#1}}            
		\put(0,0){\line(1,0){82.5}}
		\put(82.5,0){\line(0,1){#1}}
		\put(0,#1){\line(1,0){82.5}}
	\put(0,0){\unitlength=#2
	            \begin{picture}(#3)(#4)  #5
                 \end{picture}
              }
	\end{picture}
        \caption{\protect\small #6}  \medskip \hrule
     \end{figure}
	                   }

\heads{K.A. Bronnikov, M.A. Grebeniuk, V.D. Ivashchuk and V.N. Melnikov}
{Integrable Multidimensional Cosmology for Intersecting $p$-Branes}

\begin{document}
\thispagestyle{empty}
\twocolumn[
\noi \unitlength=1mm
\begin{picture}(174,8)
   \put(31,8){\shortstack[c]
       {RUSSIAN GRAVITATIONAL SOCIETY                \\
       INSTITUTE OF METROLOGICAL SERVICE             \\
       CENTER OF GRAVITATION AND FUNDAMENTAL METROLOGY}       }
\end{picture}
\begin{flushright}
                                         RGS-VNIIMS-97/08 \\
                                         gr-qc/9709006 \\
	{\it Grav. and Cosmol.} {\bf 3}, No. 2 (10), 105--112 (1997)
\end{flushright}
\bigskip

\Title{INTEGRABLE MULTIDIMENSIONAL COSMOLOGY \yy
                   FOR INTERSECTING $P$-BRANES}

\noi
{\large\bf K.A. Bronnikov\foom 1\dag, M.A. Grebeniuk\foom 2\ddag,
        V.D. Ivashchuk\foom 3\dag\ and V.N. Melnikov\foom 3\dag}
\bigskip

\noi \dag\ \ {\it Centre for Gravitation and Fundamental Metrology, VNIIMS,
          3--1 M. Ulyanovoy St., Moscow 117313, Russia}
\medskip

\noi \ddag\ \ {\it Moscow State University, Physical Faculty,
          Department of Theoretical Physics, Moscow 117234, Russia}
\medskip

{\it Received 5 April 1997}

\Abstract
{A multidimensional field model describing the behaviour of (at most) one
Einstein space of non-zero curvature and $n$ Ricci-flat internal spaces
is considered. The action contains several dilatonic scalar fields
$\varphi^a$ and antisymmetric forms $A^I$.
The problem setting covers various problems with field dependence on a
single space-time coordinate, in particular, isotropic and
anisotropic homogeneous cosmologies. When the forms are chosen to be
proportional to volume forms of ``$p$-brane" submanifolds
manifold, a Toda-like Lagrange representation arises.  Exact solutions
are obtained when the $p$-brane dimensions and the dilatonic couplings obey
some orthogonality conditions. General features and some special cases of
cosmological solutions are discussed. It is shown, in particular, that
all hyperbolic models with a 3-dimensional external space possess an
asymptotic with the external scale factor $a(t) \sim |t|$ ($t$ is the
cosmic time), while all internal scale factors and all scalar fields tend to
finite limits. For $D=11$ a family of models with one 5-brane and three
2-branes is described. }

]  
\email 1 {kb@goga.mainet.msk.su}
\email 2 {mag@gravi.phys.msu.su}
\email 3 {melnikov@fund.phys.msu.su}

\section{Introduction}

This paper continues the trend of studying various exact solutions
describing the properties of $p$-branes interacting with gravity.
Such problems naturally emerge in bosonic sectors of supergravitational
models \cite{CJS,SS} and may be of interest in the context of superstring
and M-theories [3--7].

We actually consider gravitational models containing several coupled
dilatonic scalar fields and antisymmetric forms \cite{IMO},
namely, the one-variable (in particular, cosmological)  sector
of the model from \cite{IMO}, i.e. a self-consistent set of
$D$-dimensional Einstein-Hilbert equations and
the equations for interacting dilatonic scalar
fields and fields of forms. The model describes
generalized intersecting $p$-branes (for different aspects of
$p$-branes see \cite{St,D,AIR} and references therein.) Using a
$\sigma$-model representation \cite{IMO}, we reduce the equations of motion
to a pseudo-Euclidean Toda-like Lagrange system \cite{IM3} with a
zero-energy constraint.
In the simplest case of orthogonal vectors in the exponents of the Toda
potential we obtain exact solutions.  Some general features of the
cosmological solutions and their special cases are then discussed.

Special cases of the present models were recently studied
by a number of authors (\cite{LPX,AIV} and others); thus, spherically
symmetric and cosmological models with a conformally invariant
generalization of the Maxwell field to higher dimensions were discussed
in Refs.\,\cite{Fa,BrFa,Br97}; in particular, in \cite{Br97} some integrable
cases with perfect fluid sources were indicated. The present treatment deals
with the general class of coupled $p$-brane and dilatonic fields in the
so-called electric case, but with no other material sources; the
corresponding magnetic and mixed electro-magnetic models will be discussed
elsewhere.


For convenience we give a summary of indices used in the paper and
objects they correspond to. Namely:
\begin{description}
\item[]
     $M,\ N,... \mapsto $ coordinates of the $D$-dimensional Riemannian space
	$\M$;
\item[]
     $I,\ J,... \mapsto $ subsets of the finite set $\Omega$ and
     antisymmetric forms $A^I,\ F^I$;
\item[]
	$a,\ b,... \mapsto $ scalar fields;
\item[]
	$i,\ j,... \mapsto $ subspaces of $\M$;
\item[]
	$m_i,\ n_i \mapsto $ coordinates in $M_i$;
\item[]
	$A,\ B,... \mapsto $ minisuperspace coordinates.
\end{description}
	As usual, summing over repeated indices is assumed when one of them is
	at a lower position and another at an upper one.


\section{Initial field model}

	We start (like \cite{IMO}) from the action
\bearr                                   \label{2.1}
S = \frac{1}{2\kappa^{2}}
\intl_{\M} d^{D}z \sqrt{|g|} \biggl\{ {R}[g]
     - \delta_{ab} g^{MN} \d_{M} \varphi^a \d_{N} \varphi^b \nnn
\qquad - \sum_{I\in \Omega}
 	\frac{\eta_I}{n_I!} \exp(2 \lambda_{Ia} \varphi^a) (F^I)^2
                  \biggr\}+ S_{\rm GH},
\ear
in a $D$-dimensional (pseudo-)Riemannian manifold $\M$
with the metric $g = g_{MN} dz^{M} \otimes dz^{N}$;
$|g| = |\det (g_{MN})|$;
$\varphi^a$ are dilatonic scalar fields;
\beq
F^I =  dA^I =
\frac{1}{n_I!} F^I_{M_1 \ldots M_{n_I}}
dz^{M_1} \wedge \ldots \wedge dz^{M_{n_I}}                    \label{2.2}
\eeq
are $n_I$-forms ($n_I \geq 2$);
$\lambda_{Ia}$ are coupling constants;
$\eta_I = \pm 1$ (to be specified later); furthermore,
\beq                                                          \label{2.3}
(F^I)^2 =  F^I_{M_1 \ldots M_{n_I}} F^I_{N_1 \ldots N_{n_I}}
				g^{M_1 N_1} \ldots g^{M_{n_I} N_{n_I}},
\eeq
$I \in \Omega$,  $a\in {\cal A}$,
where $\Omega$ and ${\cal A}$ are non-empty finite sets. Finally,
$S_{\rm GH}$ is the standard Gibbons-Hawking boundary term \cite{GH},
essential for a quantum treatment of the model.


The equations of motion corresponding to  (\ref{2.1}) have the form
\bearr                                                          \label{2.4}
     R_{MN} - \frac{1}{2} g_{MN} R  =  T_{MN},
                   \\ \lal
\label{2.5} \DAL \varphi^a =
	\sum_{I \in \Omega} \eta_I \frac{\lambda_{Ia}}{n_I!}
	\e^{2 \lambda_{Ib} \varphi^b} (F^I)^2,  \\ \lal
\label{2.6}
\nabla_{M}\left(\e^{2 \lambda_{Ia} \varphi^a}
				F^{I, MM_2 \ldots M_{n_I}}\right)  =  0,
\ear
where
\bearr \label{2.7}
T_{MN} =  \sum_{a\in {\cal A}} T_{MN}[\varphi^a] +
          \sum_{I \in \Omega} \eta_I
          \e^{2 \lambda_{Ia} \varphi^a} T_{MN}[F^I],\nnn \\ \lal
                                                                 \label{2.8}
T_{MN}[\varphi^a] =
              \d_{M} \varphi^a \d_{N} \varphi^a -
\frac{1}{2} g_{MN} \d_{P} \varphi^a \d^{P} \varphi^a, \\ \lal
T_{MN}[F^I] = \frac{1}{n_{I}!}  \biggl[ - \frac{1}{2} g_{MN} (F^{I})^{2}
\nnn
\cm \qquad + n_{I}
	F^{I}_{M M_2 \ldots M_{n_I}} F_{N}^{I, M_2 \ldots M_{n_I}}\biggr] .
\label{2.9}
\ear
Here $\DAL$ and $\nabla$ are the Laplace-Beltrami and covariant derivative
operators corresponding to  $g$, respectively.

We consider the manifold
\beq  \label{2.10}
	\M = \R  \times M_{0} \times \ldots \times M_{n}
\eeq
with the metric
\beq      \label{2.11}
g= w \e^{2{\gamma}(u)} du \otimes du +
\sum_{i=0}^{n} \e^{2\phi^i(u)} g^i ,
\eeq
where $w=\pm 1$, $u$ is a distinguished coordinate which, by
convention, will be called ``time";
\[
g^i  = g^i_{m_{i} n_{i}}(y_i) dy_i^{m_{i}} \otimes dy_i^{n_{i}}
\]
is a metric on
$M_{i}$  satisfying the equation
\beq   \label{2.13}
R_{m_{i}n_{i}}[g^i ] = \xi_{i} g^i_{m_{i}n_{i}},
\eeq
$m_{i},n_{i}=1,\ldots, d_{i}$ ($d_{i} = \dim M_i$);
 $\xi_i= \const$.

We assume each manifold $M_i$ to be oriented and connected,
$i = 0,\ldots,n$. Then the volume $d_i$-form
\beq    \label{2.14}
	\tau_i  = \sqrt{|g^i(y_i)|}
	\ dy_i^{1} \wedge \ldots \wedge dy_i^{d_i},
\eeq
and the signature parameter
\beq    \label{2.15}
	\eps(i)  = \sign \det (g^i_{m_{i}n_{i}}) = \pm 1
\eeq
are correctly defined for all $i=0,\ldots,n$.

Let $\Omega$ in (\ref{2.1}) be the set of all non-empty
subsets of the set of indices $\{ 0, \ldots, n \}$, i.e.
\bearr    \label{2.16}
	\Omega = \Omega(0,n) \equiv \nnn
	\{ \{ 0 \}, \{ 1 \}, \ldots, \{ n \},
		\{ 0, 1 \}, \ldots, \{ 0, 1,  \ldots, n \} \}.
\ear
The number of elements in $\Omega$ is $|\Omega| = 2^{n+1} - 1$.

For any $I = \{ i_1, \ldots, i_k \} \in \Omega$, $i_1 < \ldots < i_k$,
we put in (\ref{2.2})
\beq    \label{2.17}
	A^I = \Phi^I \tau_I
\eeq
where
\beq    \label{3.82n}
	\tau_I = \tau_{i_1}  \wedge \ldots \wedge \tau_{i_k},
\eeq
and $\tau_{i}$ are defined in (\ref{2.14}).
In components \eq(\ref{2.17}) reads:
\bearr    \label{2.18}
    A^{I}_{P_1 \ldots P_{d(I)}}(u,y) = \nnn
\quad \Phi^{I}(u) \sqrt{|g^{i_1}(y_{i_1})|} \ldots \sqrt{|g^{i_k}(y_{i_k})|}
	\ \varepsilon_{P_1 \ldots P_{d(I)}},
\ear
where $\varepsilon_{P_1 \ldots P_k}$ is the Levi-Civita symbol and
\beq    \label{2.19}
	d(I) \equiv d_{i_1} + \ldots + d_{i_k} =  \sum_{i \in I} d_i
\eeq
is the dimension of the oriented manifold
\beq  \label{2.20}
	M_{I} = M_{i_1}  \times  \ldots \times M_{i_k},
\eeq
($d(\emptyset) = 0$) and the indices $P_1, \ldots, P_{d(I)}$   correspond
to $M_I$.

It follows from (2.16) that
\beq    \label{2.21}
F^I = dA^I = d \Phi^I \wedge \tau_{I}.
\eeq
	By construction and according to (\ref{2.21}),
\beq    \label{2.23}
	n_I = d(I) + 1, \cm I \in \Omega,
\eeq
	so that the ranks of the forms $F^I$ are fixed by
	the manifold decomposition.

	For the dilatonic scalar fields we put
\beq    \label{2.24}
	\varphi^a = \varphi^a(u),\cm    a \in {\cal A}.
\eeq

	The problem setting as described embraces various classes of models
	where field variables depend on a single coordinate, such as
\begin{description}
\item[(A)]
	cosmological models, both isotropic and anisotro\-pic, where the $u$
     coordinate is timelike, $w=-1$, and some of the factor
	spaces (one in the isotropic case) are identified with the physical
	space;
\item[(B)]
	static models with various spatial symmetries (spherical, planar,
	pseudospherical, cylindrical, toroidal) where $u$ is a spatial
     coordinate, $w=+1$, and time is selected among $M_i$;
\item[(C)]
	Euclidean models with similar symmetries, or models with a Euclidean
	``external" space-time; $w=+1$.
\end{description}
     A simple analysis shows that in all Lorentzian models,
	in order to have a positive energy density
	$T^0_0$ of the fields $F^I$, one should choose in (\ref{2.1})
\beq
     \theta_I \eqdef \eta_I \eps(I)  = - w , \
     \qquad \eps(I)\eqdef \prod_{i\in I}\eps(i),
	                                                           \label{eps}
\eeq
     where, in the case B, it is assumed that $I \ni i_0$,
     where $M_{i_0}$ contains the time submanifold. Otherwise
     we simply obtain
\beq
     \theta_I = +1.
                                                     \label{ni}
\eeq

	We also use the following notations for the logarithms of
	volume factors of the subspaces of $\M$:
\beq \label{2.28}
	\sum_{i=0}^{n} d_i \phi^i \equiv \sigma_0, \cm
                 \sum_{i\in I} d_i \phi^i \equiv \sigma_I .
\eeq


\section{Sigma model representation}

As in \cite{IMO}, we pass to the sigma-model representation.
It is easy to verify that the field equations
(\ref{2.4})-(\ref{2.6}) for the field configuration
(\ref{2.11}), (\ref{2.21}), (\ref{2.24})
may be obtained as the equations of motion corresponding to the action
\bearr \label{2.29}
S_{\sigma} = \frac{1}{2 \kappa^{2}_0}
\int du \e^{\sigma_0 - \gamma} \biggl\{-wG_{ij}\dot\phi^i\dot\phi^j
	-w\delta_{ab}\dot\varphi^a\dot\varphi^b         \nnn
-\sum_{I\in\Omega}
     \e^{2\vec\lambda_I\vec\varphi -2\sigma_I} w \theta_I
          (\dot\Phi^I)^2-2V\e^{-2(\sigma_0-\gamma)}\biggr\} ,
\ear
where $\vec\varphi=(\varphi^a)$, $\vec\lambda_I=(\lambda_{Ia})$,
dots denote $d/du$,
\bearr \label{3.3}
	G_{ij}=d_i\delta_{ij}-d_id_j ,
\ear
are component of the ``pure cosmological" minisuperspace metric and
\beq  \label{3.4}
 V = {V}(\phi) = - \Half
     \sum_{i =1}^{n} \xi_i d_i \e^{-2 \phi^i + 2 {\sigma_0}(\phi)}
\eeq
is the potential.

For finite internal space volumes (e.g. compact $M_i$)
\beq   \label{2.32}
V_i = \int_{M_i} d^{d_i}y_i \sq{g^i } < + \infty,
\eeq
the action (\ref{2.29}) coincides with (\ref{2.1}) if
\beq    \label{2.34}
\kappa^{2} = \kappa^{2}_0 \prod_{i=0}^{n} V_i.
\eeq

The representation (\ref{2.29}) follows as well from a more general
$\sigma$-model action of  \cite{IMO}.

The action (\ref{2.29}) may be written in the following form corresponding to
an arbitrary time gauge:
\bearr   \label{3.5}
S_\sigma=\frac12\int du\Bigl\{(-w)
{\cal N} {\cal G}_{\hat A\hat B}(\zeta)\zeta^{\hat A}\zeta^{\hat B}
				-2{\cal N}^{-1}V(\zeta)\Bigr\} \nnn
\ear
where $(\zeta^{\hat A})=(\phi^i,\varphi^a,\Phi^{I})\in\R^{n+1+m+m'}$,
$m=|\Omega|$, $m' = |{\cal A}|$,
\bearr   \label{N}
	{\cal N}=\e^{\sigma_0-\gamma} > 0
\ear
is the lapse function and the matrix
\beq
({\cal G}_{\hat A\hat B})   \label{G}
{\arraycolsep=3.5pt
=\left(\begin{array}{ccc}
	G_{ij} &0          &0    \\[5pt]
	0      &\delta_{ab}&0 \\[5pt]
     0      &0          &\theta_I  \delta_{IJ}
                \e^{2\vec\lambda_{I} \vec\varphi-2\sigma_I}
\end{array}\right)}
\eeq
gives the minisuperspace (target space) metric.

Let us, like \cite{Br73}, choose the harmonic $u$ coordinate ($\DAL u = 0$),
such that
\bearr  \label{3.1n}
	\gamma=\sigma_0(\phi) \quad \then \quad {\cal N}=1 .
\ear
The minisupermetric ${\cal G}_{\hat A\hat B}d\zeta^{\hat
A}\otimes d\zeta^{\hat B}$ still remains to be curved.
However, the integrability problem is considerably
simplified since it is possible to integrate the generalized Maxwell
equations:
\bearr   \label{3.4n}
	\frac d{du}\left[\e^{2\vec\lambda_I\vec\varphi-2\sigma_I}
					\dot\Phi^I\right]=0 , \nnn
\label{3.5n}
	\dot\Phi^I=Q_I\e^{-2\vec\lambda_I\vec\varphi+ 2\sigma_I},
\ear
where $Q_I$ are constants.

Let
\bear   \label{3.6n}
Q_I \ne 0, \cm \lal  I \in \Omega_*,                   \nn
Q_I = 0,   \cm \lal  I \in \Omega \setminus \Omega_*,
\ear
where $\Omega_* \subset \Omega$ is a non-empty subset of $\Omega$.

For fixed $Q_I$ the Lagrange equations for
$\phi^i$ and $\varphi^a$, after substitution of (\ref{3.5n}),
are equivalent to those for the Lagrangian of a
pseudo-Euclidean Toda-like system (see \cite{IMZ,IM3,GIM}) with a
zero-energy constraint:
\bearr   \label{3.9n}
	L_Q=\frac12\oG_{AB}\dot x^A\dot x^B-V_Q , \\ \lal   \label{3.10n}
	E_Q=\frac12\oG_{AB}\dot x^A\dot x^B+V_Q=0 ,
\ear
where $x=(x^A)=(\phi^i,\varphi^a)$, $i=0,\ldots,n$; $a\in {\cal A}$,
\bearr   \label{3.11n}
(\oG _{AB})=\left(\begin{array}{cc}
  	G_{ij}&       0   \\
	0     &  \delta_{ab}
\end{array}\right)
\ear
and the potential $V_Q$ has the form
\bearr
     V_Q= \sum_{j=0}^n\left(\frac w2 \xi_i d_i\right)\e^{2U^i(x)} \nnn
\label{3.12n}
\cm   +\sum_{I\in\Omega_*} \frac{\theta_I}{2} (Q_I)^2\e^{2U^I(x)},
\ear
	where
\bearr   \label{3.13n}
     U^i(x)=U_A^i x^A=-\phi^i+ \sigma_0 , \\ \lal   \label{3.15n}
     U^I(x)=U_A^I x^A= \sigma_I -\vec\lambda_I \vec\varphi ,
\ear
	or in components
\bearr   \label{3.16n}
	(U_A^i)=(-\delta_j^i+d_j, \quad  0) , \\ \lal   \label{3.17n}
     (U_A^I)=\left(\delta_{iI}\, d_i,\quad -\lambda_{Ia}\right),
\ear
     where
\beq    \label{ind}
     \delta_{iI} \eqdef \sum_{j\in I} \delta_{ij}
\eeq
     is an indicator
     of $i$ belonging to $I$ (1 if $i\in I$ and 0 otherwise).

     The constraint (\ref{3.10n}) follows from the ${u \choose u}$
     component of the
	Einstein equations (\ref{2.4}), which contains only first-order
	derivatives with respect to $u$.

	The nondegenerate matrix $\oG_{AB}$ (\ref{3.11n}) defines an
	$(n+ 1 + m')$-dimensional
	minisuperspace metric whose contravariant components form the inverse
	matrix
\bearr   \label{3.21n}
	(\oG ^{AB})=\left(\begin{array}{cc}
		G^{ij}&      0\\
		0     &\delta^{ab}
	\end{array}\right)
\ear
	where, as in \cite{IMZ},
\bearr   \label{3.22n}
	G^{ij}=\frac{\delta^{ij}}{d_i}+\frac1{2-D}, \qquad i,j=0,\dots,n.
\ear

The integrability of the Lagrange system crucially depends on the scalar
products of the vectors $U^i$, $U^\Lambda$, $U^I$ from
(\ref{3.13n})--(\ref{3.15n}). These products are
(by definition, $(U,U') = \oG^{AB} U_A U'_B$):
\bearr   \label{3.23n}
(U^i,U^j)= \frac{\delta_{ij}}{d_j}-1 ,  \\ \lal   \label{3.24n}
(U^I,U^i)=-\delta_{iI}, \\ \lal   \label{3.27n}
(U^I,U^J)= q(I,J)+\vec\lambda_I\vec\lambda_J ,
\ear
where
\bearr   \label{3.29n}
	q(I,J)\equiv d(I\cap J) - \frac{d(I)d(J)}{D-2}.
\ear

The relation (\ref{3.23n}) was found
in \cite{IM} and (\ref{3.27n}) in \cite{IMO}
($U_A^I=-L_{AI}$ in the notations of \cite{IMO}).


\section{Solutions for at most one curved factor space}  
\subsection{Solutions with one curved factor space}      

To solve the field equations, let us adopt the following assumptions:
\begin{description}
\item[(i)]
	Among all $M_i$, only $M_0$ has a nonzero curvature, while others are
	Ricci-flat, i.e., $\xi_i=0$ for $i >0$ and (properly normalizing the
	scale factor $\phi_0$)
\beq
	\xi_0 = (d_0-1)K_0, \cm  K_0 = \pm 1.                  \label{defK0}
\eeq
\item[(ii)]
	Neither of $I \in \Omega_*$ contains the index 0, that is, neither of
	the $p$-branes under consideration involves the curved subspace $M_0$.
\item[(iii)]
	The vectors $U^I$, $I \in \Omega_*$, are mutually orthogonal with
	respect to the metric $\oG_{AB}$, that is,
\bearr                                                       \label{3.51n}
     d(I\cap J)-\frac{d(I)d(J)}{D-2}+\vec\lambda_I\vec\lambda_J=0
\ear
	for any $I\ne J$.
\end{description}

	Thus the potential $V_Q$ (\ref{3.12n}) contains one ``time-like" vector
	$U^0$ with the norm
\bearr   \label{3.31n}
	(U^0,U^0)=- (d_0-1)/d_0 < 0
\ear
	$(d_0>1)$ and $m_*=|\Omega_*|$ ``spacelike" vectors $U^I$ with the norms
	$1/N_I > 0$:
\bear
	1/N_I^2  \al \eqdef \al (U^I,U^I)           \nn       \label{3.32n}
		    \eql     d(I)\frac{D-2-d(I)}{D-2}+(\vec \lambda_I)^2>0.
\ear

	One easily verifies that under the assumptions (i)--(iii) the
	conditions adopted in the Appendix are fulfilled, with the space $V$
	of $(x^A)$ and the scalar product
     $\aver{{\bf x},{\bf y}} = \oG_{AB}x^A y^B = (x,y)$ defined in the
	previous section. Therefore we are able to write down the following
	solution to the field equations:
\bearr                                                   \label{s1}
		  x^A(u)=- \frac{d_0}{d_0-1} U^A_0 y_0(u)              \nnn
\cm +\sum_{I\in\Omega_*}N_I^2\, U_I^A\, y_i(u) + c^A u+ \oc^A,
\ear
	$A=i,a$ ($i=1,\dots,n$; $a\in {\cal A}$), where
	the functions $y_0,\ y_I$ are found from the relations
\bear
	\e^{-y_0(u)}  \eql  (d_0 -1) S (\eps_0, h_0, u-u_0), \nnn
                        \inch \cm       \eps_0 = wK_0,   \nn   \label{s3}
     \e^{-y_I(u)}  \eql  N_I^{-1} \, |Q_I|\, S (-\theta_I, h_I, u-u_I);
\ear
	the function $S(.,.,.)$ is defined in (\ref{A8});
	the constants $N_I$ in (\ref{3.32n}); recall that
	$w = \sign g_{uu}$. The integration constants
	$c^A$, $\oc^A$, $h_0$, $u_0$, $h_I$ and $u_I$ are connected by the
	following relations due to (\ref{A4}) and the energy integral
	(\ref{3.10n}):
\bear
	c^0(d_0 -1) + \sum_{i=1}^{n} d_i c^i     \eql 0; \nn   \label{s4}
	\oc^0(d_0 -1) + \sum_{i=1}^{n} d_i \oc^i \eql 0; \\
	\sum_{i\in I}
		d_i c^i - \lambda_{Ia} c^a     \eql 0; \nn          \label{s5}
	\sum_{i\in I}
		d_i \oc^i - \lambda_{Ia} \oc^a  \eql 0; \qquad I \in \Omega_*;
\ear
\vspace{-2.5mm}
\bearr                                                      \label{s6}
   \! \frac{d_0}{d_0-1} h_0^2\sign h_0
    	    = \sum_{I\in \Omega_*} N_I^2 h_I^2 \sign h_I \nnn
    \ \ + (d_0 -1) (c^0)^2
    + \sum_{j=1}^n d_j (c^j)^2 + \delta_{ab}c^a c^b.
\ear
	Finally, the contravariant components
	$U_0^{A}$, $U_I^{A}$ of the vectors $U^0$, $U^I$ are found using the
	metric tensor $\oG^{AB}$ ($U^A = \oG^{AB}U_B$) presented in
	(\ref{3.21n}), (\ref{3.22n}). Namely:
\bearr
     U_0^i= -\delta_0^i/d_i, \cm U_0^I=0 , \nnn
                                                          \label{s7}
     U_I^i=  \delta_{iI} - \frac{d(I)}{D-2},\qquad
     U_I^a= -\lambda_{Ia} ,
\ear
	$i=0,\dots,n$; $I\in\Omega_*$; $a \in {\cal A}$.

	Thus the logarithms of scale factors in the metric (\ref{2.11}) and the
	scalar fields are
\bearr                                                           \label{s8}
	\phi^i (u) = \frac{\delta_0^i}{d_0-1} y_0 (u) \nnn
      +\sum_{I\in\Omega_*} \biggl[
      \delta_{iI} - \frac{d(I)}{D-2}\biggr] N_I^2\, y_I (u)
	 								+c^i u + \oc^i,\\
\lal                                                             \label{s9}
	\varphi^a(u) = - \sumo \lambda_{Ia} N_I^2\, y_I(u) + c^a u + \oc^a
\ear
	and the  function $\gamma= \half\ln |g_{uu}| \equiv \sigma_0$  is
\bearr                                                           \label{s10}
\gamma (u) = \sum_{i=0}^n d_i\phi^i =\frac{d_0}{d_0-1}y_0 (u) \nnn
     \qquad -\sumo \frac{d(I)}{D-2}N_I^2\, y_I(u) + c^0 u + \oc^0.
\ear
Substitution of the relations (\ref{s8})--(\ref{s9}) into (\ref{3.5n})
implies
\bearr                                                         \label{s11}
	\dot\Phi^I= Q_I \, \e^{2y_I}
\ear
and hence we get for the forms
\bearr                                                         \label{s12}
     F^I=d\Phi^I\wedge \tau_I= Q_I \e^{2y_I}\, du\wedge \tau_I.
\ear

	Eqs.\,(\ref{s8})--(\ref{s12}), along with the definitions
	(\ref{s3}), (\ref{3.32n}) and (\ref{A8}) and the relations
	(\ref{s4})--(\ref{s6}) among the constants, completely determine the
	solution. Under the above assumptions (i)--(iii) the solution is
	general and involves all cases mentioned in items A, B, C at the end of
	\sect 2.

	The solution describes the behaviour of $(n+1)$ spaces
	$(M_0,g_0),\dots,(M_n,g_n)$, where $(M_0,g_0)$ is an Einstein space of
     nonzero curvature, and $(M_i,g^i)$ are ``internal" Ricci-flat spaces,
     $i=1,\dots,n$, in the presence of several scalar fields and forms.

     The solution also describes a set of charged (by the $F$ forms)
	intersecting $p$-branes ``living" on the manifolds $M_I$ (\ref{2.20}),
	$I\in\Omega_*$, where the set $\Omega_*\subset\{\{1\},\{2\},\dots,$
     $\{1,\dots,n\}\}$ does not contain $\{0\}$, i.e. all $p$-branes live in
     internal spaces.

\subsection{Solutions with Ricci-flat factor spaces}         

	The same solutions can also describe configurations where
	all $M_i$ are Ricci-flat (such as spatially flat cosmology) if one just
	deletes every mentioning of $M_0$, putting $y_0$ and the respective
	constants equal to zero and postulating the decomposition (\ref{2.10})
	in the form $\R \times M_1 \times \ldots M_n $.

	In this case \Eq{s8} for $\phi_i$ ($i=1,\ldots, n$) with $y_0 \equiv 0$
     is valid; \Eq{s9} for $\varphi^a$ is unchanged; the expression
     (\ref{s10}) for $\gamma(u)$ with $y_0\equiv 0$ is valid if we denote
\beq                                                          \label{s13}
	c^0 \eqdef \sumn d_i c^i,\qquad 	\oc^0 \eqdef \sumn d_i \oc^i.
\eeq
	The relations among the constants (\ref{s4}) disappear, while (\ref{s5})
	remain unchanged. The energy constraint may be rewritten using
	(\ref{s13}) in a simplified form:
\bearr                                                         \label{s14}
	(c^0)^2 = \sumo N_I^2 h_I^2 \sign h_I + \sumn d_i (c^i)^2
			+ \delta_{ab}c^a c^b.                       \nnn
\ear

	However, in the case of all Ricci-flat factor spaces
	there can appear forms with $d(I) = D-2$ and even $D-1$, so that zero
	or negative norms (\ref{3.32n}) are possible. Then some slight and
	evident modifications of the scheme are required; we will, however,
	always assume $d(I) < D-2$.

\medskip\noi
	{\bf Remark.} \ The above solutions evidently do not exhaust all
	integrable cases. Thus, some of the functions $y_I$ may coincide,
	reducing the number of the required orthogonality conditions
	(\ref{3.51n}) and giving more freedom in choosing the set of input
	constants $d(I)$ and $\lambda_{Ia}$. Such a coincidence of, say,
	$y_I$ and $y_K$ is a constraint on the unknowns and the emerging
	consistency relation should lead to a reduction of the number of
	integration constants (e.g. coincidence of charges $Q_I$ and $Q_K$). In
	other words, it becomes possible to obtain less general solutions but
	for a more general set of input parameters. An example of such a
	situation is given in Ref.\,\cite{Pd97} discussing some spherically
	symmetric solutions with intersecting electric and magnetic $p$-branes.


\section{Cosmological solutions}

\subsection {General features}

	Among the above solutions, cosmological models correspond to
     $w = -1$. We also assume that the curved subspace $M_0$ ($K_0= \pm 1$)
	is the ``external" physical space, spherical ($K=+1$) or hyperbolic
	($K=-1$). To describe spatially flat models (to be labelled $K_0=0$)
	we assume that in the solution of Subsec. 4.2 (where $M_0$ is absent),
	one of the subspaces, say, $M_1$, is external and is not involved in
	the $p$-branes, i.e., $ 1 \not\in I,\ I\in \Omega_*$.

     Due to $w=-1$ and positive energy condition (\ref{eps})
     we immediately obtain
\beq
     \e^{-y_I} = \frac{|Q_I|}{h_I N_I} \cosh [h_I(u-u_I)],\qquad
				 h_I > 0.                                 \label{5.1}
\eeq
	Another sign factor in (\ref{s3}), $\eps_0$, is $\eps_0 = -K_0$.

	The behaviour of the solutions is extremely diverse according to the
	interplay of the numerous integration constants and input parameters.
	Thus, different scale factors may be monotone or not, grow to infinity
	or fall to zero, or tend to constant values. Some general observations
	can nevertheless be made.

	The first observation is that the range of the time coordinate $u$ is
	$\R$ for $K_0 = 0,\ +1$. At both ends of the evolution, $u\to \pm
	\infty$, the scale factors $\e^{\phi_i}$ and the function $g_{uu} =
	\e^{2\gamma}$ behave like
\beq
	f(u) = (\cosh u)^k \e^{cu}                              \label{5.2}
\eeq
	where the constants $k$ and $c$ are fixed for all $\e^{\phi_i}$ and
	$\e^{\gamma}$ for each specific set of the input and integration
	constants.

	For spatially flat models, $K_0 = 0$, the constant $k$ is positive for
	both the ``physical" scale factor $a(u) = \e^{\phi_1(u)}$ and
     $\e^{\gamma}$. This means that $a(u) \to \infty$ at least one of the
	asymptotics and so does the cosmic synchronous time
	$t = \int \e^{\gamma} du$ (but not necessarily at the same
	asymptotic).

	For closed models ($K_0 =+1$) even such small general information cannot
	be extracted: both $a(u)$ and $\e^{\gamma}$ behave according to
	(\ref{5.2}), but now both constants $k$ and $c$ can have arbitrary
	signs.

	One can note that $f$ can have a finite limit at one end
	of the evolution (if $c= \pm k$), but in the generic case all
	$\e^{\phi_i(u)}$ and $t(u)$ have exponential asymptotics with
	different increments, hence the dependence $\e^{\phi_i (t)}$ has a
	power-law asymptotic, $\e^{\phi(u)} \sim t^s$. In particular, one can
	expect $s \gg 1$ for some choices of the parameters, i.e., power-law
	inflation in at least some spatial directions.

	For hyperbolic models, $K_0= -1$, we have in (\ref{s3}) $\eps_0=+1$
	and due to (\ref{s6}) $h_0 > 0$, so that, with no generality loss, we
	can write
\beq
	\e^{-y_0} = \frac{d_0 -1}{h_0} \sinh h_0 u          \label{5.3}
\eeq
	and $ u>0 $. As $u \to \infty $, all $\e^{\beta_i}$ and $\e^{\gamma}$
	behave again like $f$ in (\ref{5.2}), with all the above inferences,
	but, as $u\to 0$, all factors like $\e^{cu}$ and $\cosh k(u-u_0)$ are
	finite and the model behaviour is governed by $y_0$. In particular, for
	the physical dimension $d_0 =3$ we obtain:
\bearr
	\e^{\gamma} \sim u^{-3/2}, \nnn
	a(u) = \e^{\phi_0} \sim 1/\sqrt{u}, \qquad
	\qquad t\sim 1/\sqrt{u}.                              \label{5.4}
\ear
	This asymptotic corresponds to infinite linear expansion or
	contraction in the external space ($a(t) \sim |t|$), while all internal
	scale factors and scalar fields tend to finite limits.

\subsection{Some special solutions}                    

Consider the case $D=11$, $\vec\lambda_I=0$,
and $d_0=3$ (a curved isotropic physical space).
From (\ref{3.51n}) we have the following possibilities:
\beq                                                    \label{3.86n}
	\{d(I),d(J)\}=\{3,3\},\{3,6\},\{6,6\}
\eeq
if $d(I\cap J)=1,2,4$, respectively, and $2N_I^2=1$, $I,J\in\Omega_*$.
Using the rules (\ref{3.86n}), we can list all possible sets $\Omega_*$,
or collections of intersecting 2-brane and 5-branes. Two variants of maximum
possible sets $\Omega_*$ in the remaining 7 internal dimensions
are schematically shown in Fig.\,1, where
straight lines and circles correspond to possible sets of indices $I$ and
black spots refer to coordinates in the subspaces $M_i$, $i>0$.

\linethickness{0.4pt}

\Picture{60}{0.61mm}{120,55}{0,-15}
{
\put(20.00,20.00){\line(3,5){20.33}}
\put(41.00,55.00){\line(1,-2){17.67}}
\put(58.00,20.00){\line(-1,0){38.00}}
\put(20.00,20.00){\line(5,3){30.00}}
\put(31.00,38.00){\line(3,-2){27.00}}
\put(86.00,32.00){\line(1,0){24.00}}
\put(92.00,42.00){\line(3,-5){12.67}}
\put(104.00,42.00){\line(-3,-5){12.67}}
\put(16.00,21.00){\makebox(0,0)[cc]{1}}
\put(26.00,38.00){\makebox(0,0)[cc]{2}}
\put(37.00,55.00){\makebox(0,0)[cc]{3}}
\put(54.00,37.00){\makebox(0,0)[cc]{4}}
\put(62.00,20.00){\makebox(0,0)[cc]{5}}
\put(40.00,15.00){\makebox(0,0)[cc]{6}}
\put(39.00,36.00){\makebox(0,0)[cc]{7}}
\put(88.00,17.00){\makebox(0,0)[cc]{1}}
\put(81.00,31.00){\makebox(0,0)[cc]{2}}
\put(89.00,44.00){\makebox(0,0)[cc]{3}}
\put(108.00,44.00){\makebox(0,0)[cc]{4}}
\put(114.00,31.00){\makebox(0,0)[cc]{5}}
\put(108.00,17.00){\makebox(0,0)[cc]{6}}
\put(98.00,36.00){\makebox(0,0)[cc]{7}}
\put(98.00,32.00){\circle*{2.00}}
\put(105.00,21.00){\circle*{2.00}}
\put(91.00,21.00){\circle*{2.00}}
\put(85.00,32.00){\circle*{2.00}}
\put(110.00,32.00){\circle*{2.00}}
\put(104.00,42.00){\circle*{2.00}}
\put(92.00,42.00){\circle*{2.00}}
\put(41.00,55.00){\circle*{2.00}}
\put(31.00,38.00){\circle*{2.00}}
\put(50.00,38.00){\circle*{2.00}}
\put(58.00,20.00){\circle*{2.00}}
\put(20.00,20.00){\circle*{2.00}}
\put(41.00,14.00){\line(0,1){1.00}}
\put(41.00,55.00){\line(0,-1){35.00}}
\put(41.00,32.00){\circle*{2.00}}
\put(41.00,20.00){\circle*{2.00}}
\bezier{124}(31.00,38.00)(40.50,49.00)(50.00,38.00)
\bezier{124}(31.00,38.00)(24.00,24.00)(41.00,20.00)
\bezier{112}(41.00,20.00)(55.00,22.50)(50.00,37.00)
\bezier{56}(92.00,42.00)(98.00,45.00)(104.00,42.00)
\bezier{56}(104.00,42.00)(109.00,40.00)(110.00,32.00)
\bezier{56}(110.00,32.00)(110.00,24.00)(105.00,21.00)
\bezier{64}(105.00,21.00)(98.00,17.00)(91.00,21.00)
\bezier{60}(91.00,21.00)(86.00,25.00)(85.00,32.00)
\bezier{60}(85.00,32.00)(86.00,38.50)(92.00,42.00)
\put(41.00,0.00){\makebox(0,0)[cc]{\bf a}}
\put(98.00,0.00){\makebox(0,0)[cc]{\bf b}}
   }
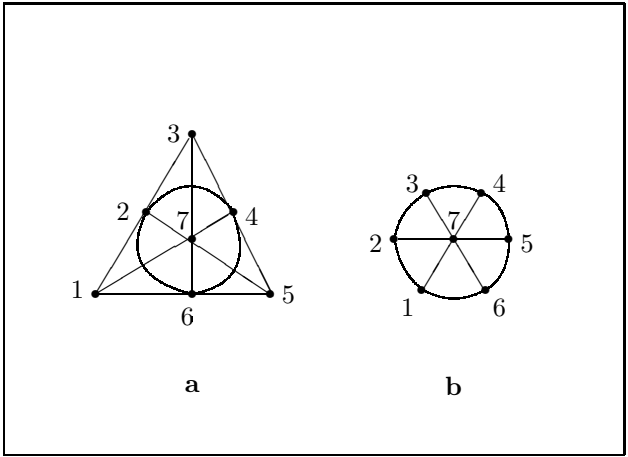
{Diagram of maximal sets of intersecting 2-branes.
Each $I$ is shown by a straight line or a circle.
Enumerated black spots correspond to internal coordinates.
\protect\medskip
\protect\newline
{\bf a:} Seven 3-dimensional sets $I$.
The subspaces $M_i$ ($i=1,\ldots,7$) are 1-dimensional.
\protect\medskip
\protect\newline
{\bf b:} Three 2-branes and one 5-brane.
The subspaces $M_i$ ($i=1,2,3,4$) are described by the coordinates labelled
as $\{1,4\}$, $\{2,5\}$, $\{3,6\}$ and $\{7\}$, respectively.
The 3-dimensional sets $I=B,C,D$ are shown by straight lines and the
6-dimensional one, $A$, by a circle.
}

     Fig.\,1a shows a system of $n=7$ one-dimensional spaces $M_1$, so that
	each black spot depicts both a coordinate and a factor space. Each of
	the 7 sets $I \in \Omega_*$ contains 3 elements, i.e. corresponds to a
	2-brane.

     Fig.\,1b shows a system of 2-branes and a 5-brane with $n=4$:
	$d_0=3$, $d_1=d_2=d_3=2$, $d_4=1$.
	Let us consider this case in more detail. The set $\Omega_*$ reads:
\beq   											\label{03}
	\Omega_*=
     \{\ubr{\{1,2,3\}}_A,\
     \ubr{\{1,4\}}_B,\ \ubr{\{2,4\}}_C,\ \ubr{\{3,4\}}_D\}
\eeq
	where the letters $A,B,C,D$ label the corresponding sets
	$I\in \Omega_*$.

	Due to the orthogonality relations between $c^i$ and $\oc_i$ all these
	constants vanish and
\beq
     3 h_0^2  = h_A^2 + h_B^2 + h_C^2 + h_D^2.             \label{05}
\eeq
	One easily finds that
\bear
     2\phi^0(u) \eql y_0 - \Sigma, \nn
     2\phi^1(u) \eql y_A + y_B - \Sigma, \nn
     2\phi^2(u) \eql y_A + y_C - \Sigma, \nn
     2\phi^3(u) \eql y_A + y_D - \Sigma, \nn
     2\phi^4(u) \eql y_B + y_C +y_D - \Sigma, \nn
     2\gamma(u)  \eql 3y_0  - \Sigma,                  \label{06}
\ear
	where all $y(u)$ are defined in (\ref{s3}) and
\beq
   \nhq  \Sigma \eqdef \sumo\! \frac{d(I)}{D-2} y_I(u)
	 = \frac{2}{3} y_A + \frac{1}{3} (y_B + y_C + y_D).   \label{07}
\eeq

	If we assume that $M_0=S^{3}$, all $y(u)$ are expressed in
	terms of hyperbolic cosines. Their asymptotics as $u \to \pm \infty$ are
\beq
	y_0 \sim -h_0 |u|, \cm  y_I \sim -h_I u.             \label{08}
\eeq
	On the other hand, the logarithms of volume factors of extra dimensions
     ($\sigma_{\rm extra}$) and the whole space ($\sigma_0$) are
\beq
	\sigma_{\rm extra}= \sum_{i=1}^{4} d_i\phi^i, \qquad
	\sigma_0 = \gamma = \sum_{i=0}^{4} d_i\phi^i.         \label{09}
\eeq
	From (\ref{06})--(\ref{08}) it follows that, as $u \to \pm\infty$,
	the volume factor of extra dimensions
\beq
	V_{\rm extra} = \e^{\sigma_{\rm extra}} \to 0         \label{010}
\eeq
	and, in addition, in view of (\ref{05}),
\beq
	V_{\rm total} = \e^{\sigma_0} = \e^{\gamma}\to 0.
\eeq
	On the other hand, as $\gamma \sim -h_0 |u|$ as $u \to \pm\infty$,
	the integral $t = \int \e^{\gamma}du $ converges in the same limits.
	One has to conclude that in these models both the total volume and that
	of extra dimensions collapse to zero at some finite times $t$ both in
	the past and in the future. The physical space, taken separately, may,
	however, avoid the collapse if one properly chooses the values of
     $h_I$ (see the first line of (\ref{06})).

	All this is true as well for hyperbolic models ($K_0=-1$), but only at
	one end of the evolution ($u\to \infty$), while at the other ($u\to 0$)
	the asymptotic is always as described at the end of Subsec.\,5.1.

	The outlined picture is true for the model with the maximal $\Omega_*$
	given in (\ref{03}); non-maximal models, with $\Omega_*$ taken as its
	subsets, require a separate consideration.

\section{Appendix}
\renewcommand{\theequation}{A.\arabic{equation}}
\sequ{0}
\def\x{{\bf x}}
\def\b{{\bf b}}
\def\c{{\bf c}}

	Let there be an $n$-dimensional real vector space $V$ with a
	nondegenerate real-valued quadratic form $\aver{.,.}$. Let, further, a
	real-valued vector function $\x(t) \in V$ obey the equations of motion
	corresponding to the Lagrangian
\bearr   \label{A1}
	L=\aver{\dot\x,\dot\x}-\sum_{s=1}^m A_s\e^{2\aver{\b_\ss,\x}},
\ear
	where $m\leq n$, $A_s = \const\ne 0$ and $\b_s \in V$ are constant
	vectors such that
\beq                                                         \label{A2}
	\aver{\b_s,\b_s} \ne 0, \cm \aver{\b_s,\b_p}=0, \quad s\ne p.
\eeq
	Then, the equations of motion for the Lagrangian (\ref{A1}) have
	the following solutions \cite{GIM}:
\bearr                                                        \label{A3}
     \x(t)= \sum_{s=1}^m\frac{\b_s}{\aver{\b_s,\b_s}} y_s(t)
         +t\c +\c_0,
\ear
	where $\c,\c_0\in V$ (integration constants) satisfy the
	orthogonality relations
\bearr   \label{A4}
	\aver{\c,\b_s} = \aver{\c_0,\b_s}=0 ,
\ear
	and the functions $y_s(t) \equiv \aver{\b_s,\x}$ are found from the
	decoupled equations
\beq
	\ddot y_s = -A_s \aver{\b_s,\b_s} \e^{2y_\ss}.        \label{A5}
\eeq
	Their solutions may be in turn presented in the form
\beq                                                         \label{A6}
	\e^{-2y_\ss(t)} = |A_s \aver{\b_s,\b_s}| S^2(\eps_s,\ h_s,\ t-t_s),
\eeq
	where $h_s,\ t_s$ are integration constants,
\beq
	\eps_s = - \sign [A_s \aver{\b_s,\b_s}]                \label{A7}
\eeq
	and the function $S(.,.,.)$ is, by definition,
\bear
	S(1,\ h,\ t) \eql \vars{ h^{-1} \sinh ht, \quad & h>0,\\
				                        t,       & h=0,\\
					     h^{-1} \sin ht,     & h<0; }\nn
     S(-1,\ h,\ t)\eql h^{-1} \cosh ht; \cm  h> 0.
                                                              \label{A8}
\ear
    The conserved energy corresponding to the solution (\ref{A3}) is
\bear                                                       \label{A9}
    E \eql \aver{\dot\x,\dot\x} + \sum_{s=1}^m A_s \e^{2y_\ss}\nn
      \eql \sum_{s=1}^m \frac {h_s^2\sign h_s}{\aver{\b_s,\b_s}}
	           +\aver{\c,\c}.
\ear

\Acknow
{This work was supported in part  by DFG grants
436 RUS 113/7, 436 RUS 113/236/O(R),
the Russian State Committee for
Science and Technology, and Russian Basic Research Foundation,
project N 95-02-05785-a.}

\end{document}